\documentclass[manuscript]{revtex4}  
\usepackage{color,graphicx}
\usepackage{amssymb,amsmath}

\begin{document}                

\title{Faddeev calculations of break-up reactions with realistic experimental constraints }

\author{J. Kuro\'s-\.Zo{\l}nierczuk$^1$, P. Th\"orngren Engblom$^2$, 
H-O.~Meyer$^1$, T.~J.~Whitaker$^1$,  H.~Wita{\l}a$^3$, 
J.~Golak$^{3}$, H.~Kamada$^4$, A.~Nogga$^5$, R.~Skibi\'nski$^3$}

\affiliation{$^1$Department of Physics, Indiana University, Bloomington, IN 47405, USA} 
\affiliation{$^2$Department of Radiation Sciences, Uppsala University, 
         S - 75121 Uppsala,  Sweden   }
\affiliation{$^3$M. Smoluchowski Institute of Physics, Jagiellonian University,
          Reymonta 4,  30-059 Krak\'ow, Poland}
\affiliation{$^4$Department of Physics, Faculty of Engineering,\\ Kyushu Institute
          of Technology,   Kitakyushu 804-8550, Japan}
\affiliation{$^5$ Institute for Nuclear Theory, University of Washington, 
Box  351550, Seattle, WA 98195, USA
}


\begin{abstract}                
We present a method to integrate predictions from a theoretical model of 
a reaction with three bodies in the final state over the region of phase 
space covered by a given experiment. 
The method takes into account the true experimental acceptance, as well as 
variations of detector efficiency, and eliminates the need for a Monte-Carlo 
simulation of the detector setup. The method is applicable to kinematically
complete experiments. Examples for the use of this method include 
several polarization observables in $dp$ break-up at 270 MeV. The calculations 
are carried out in the Faddeev framework with the CD Bonn nucleon-nucleon 
interaction, with or without the inclusion of 
an additional three-nucleon force.
\end{abstract}

\maketitle

\section{ INTRODUCTION}
For nuclear reactions with two particles of given masses in the exit 
channel, two parameters are required to specify the final state. Usually, 
these are taken to be the polar  angle and the azimuth  of one of the 
particles. The direction of the other particle and both energies 
are then given 
by energy and momentum conservation.

When three particles are present in the final state, one more momentum 
vector needs to be specified, thus three more parameters are required. 
This means that all observables are a function of five 
parameters. Often these are taken to be the polar angles and the    
azimuths of two of the three particles and one parameter that describes 
how the kinetic energy is shared by the three particles. 
This parameter could be, e.g., the  relative energy 
of  particles 1 and 2.

In order to present and discuss experimental results, 
one must select one or two 
independent variables.   
Traditionally, two outgoing particles are measured in coincidence by two 
(small) detectors. The detector positions determine the polar 
and azimuthal angles of 
both particles. In a plot of the energy of the first versus the energy 
of the second particle, all events lie on a locus. 
The position along that locus
then serves as the independent variable. The choice of the detector angles is 
arbitrary, but their solid angles must be reasonably small to keep the locus 
defined. To explore the full kinematics of the reaction the experiment must 
be repeated with different detector positions
(see e.g.~\cite{old-exp}).

Modern nuclear detection techniques make it possible to build experiments
that simultaneously cover a sizable fraction of the entire phase space of 
the reaction. 
Measuring the momenta of two outgoing particles yields enough 
information to define the complete kinematics of an event, 
with one redundant parameter 
to spare that may be used for event identification. One thus has the  
freedom of  choosing any independent variable by sorting the measured events
 accordingly.

When partitioning the phase space with respect to a given single variable, 
all other kinematic parameters are ignored. When one wants to compare the 
experiment to a model, the corresponding calculation must integrate over 
these ignored parameters. 
To carry out this integral, one  must take into account the boundaries 
of the acceptance of the experiment, as well as variations of the detection 
efficiency within the acceptance. It is usually quite difficult to describe 
the detailed performance of the detection system, and to implement 
this information in the calculation of an observable by   
a theoretical method. In this paper we present 
a straight-forward method to accomplish this task.

In Sec.~\ref{sec:method} we describe the proposed method. 
In Sec.~\ref{sec:theory} 
we outline the underlying theory of three-nucleon break-up reactions and
discuss how the calculations can be accelerated for the use in the present 
context.
In Sec.~\ref{sec:examples} the new method is applied to the analysis of 
several polarization observables in the $dp$ break-up reaction. These 
calculations are based on the charge-dependent (CD) Bonn 
nucleon-nucleon (NN) potential~\cite{cdbonn1,cdbonn2}. We also study the 
effect of including the modified Tucson-Melbourne three-nucleon force 
(TM' 3NF)~\cite{TM'-1,TM'-2}. Finally, we summarize our results in 
Sec.~\ref{sec:summary}.

\section{The Sampling Method}
\label{sec:method}

Let us denote by $x = \{\alpha_1,..,\alpha_m ,...\}$ 
the set of parameters that is needed to completely 
describe the kinematics of a given nuclear reaction. 
To specify the phase-space coordinates of a three-body-final 
state one requires 5 parameters.
  
The differential cross section for the reaction 
with unpolarized collision partners is 
given by $\sigma_0(x)$. A typical polarization observable 
$O(x)$ has the effect of modifying the 
unpolarized cross section such that 
$\sigma(x) = \sigma_0(x) (1+PO(x))$, where $P$ is the polarization 
of the beam or the target or their product, 
and $O(x)$ is a beam analyzing power, a target 
analyzing power or a spin correlation coefficient. 
In order to measure $O(x)$ one carries out 
two measurements with opposite sign of the polarization $P$. 
The yields $N_{\pm}$ accumulated during the two 
measurements in a phase space region $(x, x+\Delta x)$ are then given by
\begin{equation}
 N_{\pm}(x)=
L_{\pm} \epsilon (x) \sigma_0(x) (1\pm P _{\pm} O(x))
\label{eq:method-1} 
\end{equation}
Here, $P_+$ and $P_-$ are the magnitudes 
of the two polarizations with opposite sign, and $L_+$ and $L_-$ are 
the time-integrated luminosities for the two measurements. 
The detector efficiency $\epsilon (x)$ measures the  probability with which 
an event of interest gets registered by the detector system. For events outside 
the acceptance of the detector, $\epsilon (x) = 0$. 

Let us assume, for the moment, that the integrated 
luminosities and polarizations for the two 
measurements are the same, 
or $L_+ = L_- = L/2$, and $P_+ = P_-= P$. 
We then find for the total number of collected events
\begin{equation}
  \label{eq:method-2}
  N(x)\equiv N_+ + N_- =L \epsilon (x) \sigma_0 (x)
\end{equation}
and for the polarization observable in terms of the measured yields

\begin{equation}
  \label{eq:method-3}
  O(x)=(1/P)(N_+ - N_-)/ (N_+ + N_-)
\end{equation}

The equation for $O$ holds for any point 
$x$ in phase space, but it is obviously impractical to 
evaluate or discuss observables 
in a five-dimensional parameter space. Rather, 
one may select a single independent parameter $\alpha_m$,
ignoring all others. When the experiment ignores a given parameter, 
the full range of that parameter within the detector acceptance is 
included. Thus, for each 
of the $\alpha_m$ bins, one actually evaluates $O(\gamma )$ in a region 
$\gamma  $ of phase space, where $\gamma $ denotes the range for 
each of the ignored parameters. Assume now that we wish to 
compare an experiment $O(\gamma )$ to a theoretical model. The model 
calculation provides us with a value $O^{th}(x)$ at any point $x $ in phase 
space. In order to obtain the theoretical equivalent $O^{th}(\gamma )$ 
of the experiment we have to average $O^{th}(x)$, over the region $\gamma $,
weighted by the unpolarized cross section times the experimental 
efficiency, 

\begin{equation}
  \label{eq:method-4}
  O^{th}(\gamma )= \frac{\int_{\gamma } \sigma_0 (x) \epsilon (x) O^{th}(x) dx}
{\int_{\gamma } \sigma_0 (x) \epsilon (x) dx}.
\end{equation}

We determine $\epsilon (x) \sigma_0(x)$ by making use of Eq.~(\ref{eq:method-2}),
 and we replace the integrals in Eq.~(\ref{eq:method-4}) by 
sums over all elements $x_i$ of size $\Delta x$, that make up the region 
$\gamma $,
\begin{equation}
  \label{eq:method-5}
  O^{th}(\gamma )=\frac{\sum N(x_i) O^{th}(x_i)}{\sum N(x_i)}
\end{equation}
Here, $N(x_i)$ is the number of events collected in element $x_i$, 
irrespective of polarization. Since we are free to choose the size 
$\Delta x$ of the element, we decrease it until all $N(x_i)$ are either 
0 or 1. The number of occupied $x $ elements then equals 
the total number of events 
$N(\gamma )$ collected in region $\gamma $ during the experiment, and 
the list of the $x_i$'s for occupied bins is then identical to 
the list of phase space coordinates $x_k$ ($k = 1.. N(\gamma )$) 
for all collected events. For a kinematically complete experiment, 
in which the phase space coordinates $x_k$ are 
known for each event, the correctly averaged value for 
a calculation is thus obtained as the mean of 
the corresponding theoretical values  $O^{th}$ for all events $k$
\begin{equation}
  \label{eq:method-6}
  O^{th}(\gamma )=<O^{th}>=\frac{\sum  O^{th}(x_k)}{ N(\gamma )}
\end{equation}
This simple recipe constitutes our proposed method: in order to obtain the 
average theoretical value, correctly weighted by the product 
of unpolarized cross 
section and detector efficiency, one determines the theoretical 
$O^{th}$ for each collected event, sums these values and divides by 
the number of events. 
 
It is easy to see, that the standard deviation of the theoretical 
value that arises from the 
randomness of the experimental phase space points that are used 
to sample the region $\gamma $ is given by 
\begin{equation}
  \label{eq:method-7}
 \delta  O^{th}(\gamma )=
\sqrt{\frac{<{O^{th}}^2> - {<O^{th}>}^2 }{N(\gamma ) - 1}}
\end{equation}

The method of Eq.~(\ref{eq:method-6}), in effect, involves an average over 
theoretical values, weighted by the density in phase space of the actual 
events. This density is proportional to the unpolarized cross section and 
detector efficiency, but it also depends on the polarization asymmetry. 
The latter cancels only if the magnitudes  of the two 
polarizations with opposite sign are the same, and the corresponding, 
time-integrated 
luminosities $L_+$ and $L_-$ are the same, or $\Delta P \equiv (P_+ - P_-)/2$
and $\Delta L \equiv (L_+ - L_-)/2$ both vanish. If $\Delta P \neq 0$, then the 
averages $<O^{th}>^+$ and $<O^{th}>^-$, taken with just the data points 
with positive or negative polarization, respectively, are different. It is 
easy to see that in this case, the desired theoretical average $<O^{th}>_0$
that is free of polarization effects can be obtained from the average
$<O^{th}>$ obtained with all events according to Eq.~(\ref{eq:method-6}), 
by subtracting a correction term,
\begin{equation}
  \label{eq:method-8}
 <O^{th}>_0=<O^{th}> - \left( \frac{\Delta P}{P} - \frac{\Delta L}{L} \right)
\frac{<O^{th}>^+ - <O^{th}>^-}{2}
\end{equation} 
In the experiment~\cite{IUCFexp} to which we later apply our method, 
an effort was made 
to keep $\Delta P$ and $\Delta L$ small, and the correction term 
of~Eq.~\ref{eq:method-8} turned out to be insignificant.

\section{Theory used to calculate observables}
\label{sec:theory}
\subsection{Basic definitions} 

As an example of an application of the method 
described in this paper, let us consider
 the proton-deuteron break-up reaction at an energy below 
the pion production threshold. Here, all three nucleons are moving freely 
in the outgoing channel.
The kinematics of a three body final state is determined 
by nine variables, but energy and momentum conservation 
reduces the number of independent variables to five.

\begin{figure}
[htbp]
  \centering
  \includegraphics[width=8cm,bb=0 0 360 200,clip=true]{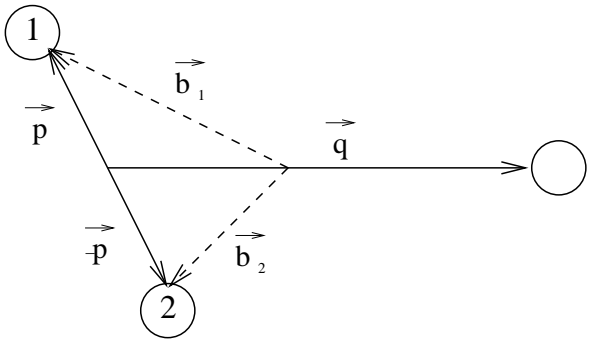}
\caption{The momenta of three particles in center-of-mass system. 
Particles numbered 1 and 2 are the two protons 
with momentum $\vec b_1$ and $\vec b_2$.} 
\label{fig:3n}
\end{figure}

Assume that a $d+p \rightarrow p+p+n$   is kinematically complete, 
as is the case when the
energies and directions of the two final-state protons are detected.
To describe the final state kinematics we define the Jacobi 
momenta $\vec p$ and $\vec q$:
\begin{eqnarray}
\vec p&\equiv &\frac{1}{2}(\vec b_1-\vec b_2)\nonumber \\
\vec q&\equiv &-(\vec b_1+\vec b_2)
\label{eq:jacoby}
\end{eqnarray}
where $\vec b_1$ and $\vec b_2$ are momenta of the two protons in 
the center-of-mass system (see Fig.~\ref{fig:3n}). 

In the remainder of this paper we are concerned only with 'axial' 
observables that are symmetric with respect to a rotation around the beam axis.
For this special case, the number of independent variables reduces to four, 
since only the difference between the azimuths is relevant. We choose these 
variables to be 
$x = \{ p, \theta_p, \theta_q, \Delta\phi  \equiv   \phi_p -\phi_q \}$, 
in the center of mass.

\subsection{The break-up amplitude}
\label{sec:Faddeev}
The theoretical predictions presented in this work  are based 
on solutions of 3N Faddeev equations 
using realistic NN interaction and a $2\pi$-exchange   
three-nucleon force model. 
In the following, we give a short overview of the underlying formalism. 
For more details we refer to~\cite{raport,huber97} and references therein. 

The transition operator for the break-up process $U_o$ 
can be expressed in terms of a $T$ amplitude as:
\begin{equation}
U_o=(1+{\cal P})T
\label{eqU}
\end{equation}
The operator $T$ fulfills Faddeev-like equation
\begin{equation}
T = t{\cal P}+( 1 + t G_0 )  V_4^{(1)}  ( 1 + {\cal P} ) + t{\cal P}G_0T 
+  ( 1 + t G_0 ) V_4^{(1)} ( 1 + {\cal P} ) G_0  T
\label{eqT}
\end{equation}
where the two-body  $t$-operator is denoted by $t$, 
the free 3N propagator by $G_0$ and $\cal P$ is 
the sum of a cyclical and an anti-cyclical 
permutation of three particles. 
The three-nucleon  force $V_4$ can always be decomposed into 
a sum of three parts:
\begin{equation}
\label{eqV4}
V_4 = V_4^{(1)} + V_4^{(2)} + V_4^{(3)} ,
\end{equation}
where $V_4^{(i)}$ singles out nucleon $i$ and 
 is symmetrical under the exchange of the  other two ones.
As seen in Eq.~(\ref{eqT}) only 
one of the three parts occurs explicitly, the others enters 
via the permutations
contained in $\cal P$.

As a 2N-force we use the CD Bonn NN 
potential~\cite{cdbonn1,cdbonn2}. This interaction is one of the modern, 
phenomenological models which describe the 2N data set with a $\chi^2$ 
per data point that is very close to one. As a 3NF we take the new version of 
the Tucson-Melbourne force - TM' 3NF~\cite{TM'-1, TM'-2}. The strong 
cutoff parameter $\Lambda $ has been adjusted such that the 3NF together
with CD Bonn potential reproduces the measured triton binding 
energy ($\Lambda=4.593m_{\pi}$~\cite{TM'-2}).

We solve Eq.~(\ref{eqT}) in a partial-wave projected momentum-space 
basis~\cite{raport}. The calculations shown here are carried out for a
deuteron bombarding energy of $E_d=$270 MeV (the equivalent laboratory energy 
of incoming proton is $E_p=135$ MeV). In order to achieve converged solutions 
of the Faddeev equations, it is necessary to include partial waves up to the 
2N-subsystem total angular momentum $j_{max}=5$.
This corresponds to up to $142$ partial wave states in the 3N system. 
The 3NF includes all total angular momenta of the 3N system 
up to $J=13/2$ while 
the longer-ranged 2N interactions require states up to  
$J=25/2$ for convergence.
The Coulomb force between the protons is neglected in these calculations.
Coulomb effects are expected to be small at the considered energy of the
incoming proton and to show up only in final-state 
configurations with low relative
proton momenta.

The Faddeev calculation results in the break-up amplitude $U_o $. From this 
amplitude one obtains, in a standard manner, the cross section or any 
polarization observable~\cite{raport}. 
Definitions of polarization observables with a spin-1 
and a spin-1/2 particle in the initial state can be found in~\cite{ohlsen72}.

In the following, we will discuss the longitudinal proton analyzing power 
$A_z$ which is non-zero for non-coplanar $\vec p$ and $\vec q$, but forbidden
by parity conservation when $\Delta\phi$ equals 0 or $\pi $. 
In addition we have
calculated the deuteron tensor analyzing power $A_{zz}$
 and some spin correlation coefficients $C_{a ,b }$,      
that can be measured with both initial-state particles polarized. Here, 
$a $ refers to the vector or tensor polarization of the deuteron,  and 
$b $ to the polarization direction of the nucleon.

\subsection{Computational details}
\label{sec:Computational details}
In Sec.~\ref{sec:method} we have presented a method to calculate the 
theoretical estimate for any observable, correctly weighted by the 
experimental acceptance and detector efficiency. The method requires that a
theoretical value is obtained for the phase space coordinates of every 
measured event. However, deducing such a value directly from the Faddeev 
amplitude (single-shot mode) for typically 
a few million events is not practical.
 In order to overcome this limitation, we precalculate and 
store the desired observable ($O^{th}(x)$) at discrete points, 
covering the entire phase space. The value of the observable at any phase 
space point is then retrieved by multidimensional 
linear interpolation~\cite{bib:LinInt} 
as described in the Appendix.

The observable is precalculated at uniformly-spaced phase-space points,
called a 'grid', chosen as follows. The azimuth variable $\Delta \phi$ is 
taken in steps of $10^\circ$ between 0 and $360^\circ$. The polar angles 
$\theta_p$ and $\theta_q$ are taken in $5^\circ$ steps, 
from $5^\circ$ to $90^\circ$
 and $180^\circ$, 
respectively. For $\theta_p$ only half the angular range is needed 
because the two observed particles in the final state are identical. 
The range of allowed $p$ values is divided into 30 steps. Thus, the grid 
points form a matrix with 719,280 elements. 

\begin{figure}
  \centering
  \includegraphics[width=10cm,bb=0 0 730 530,clip=true]{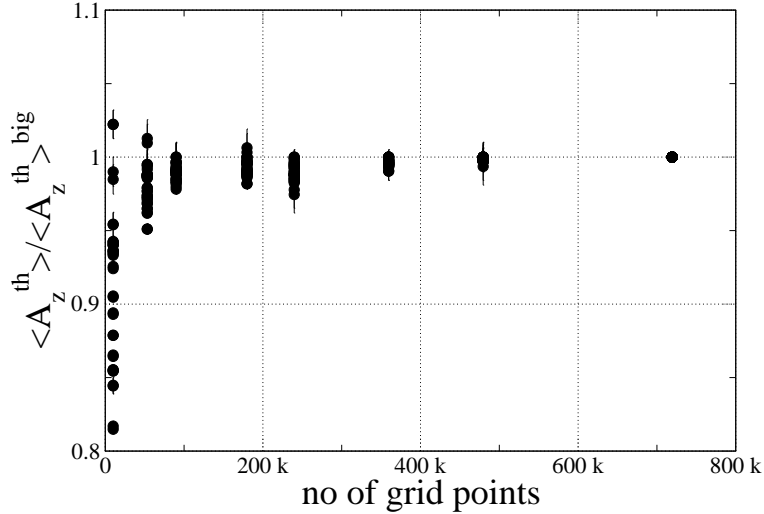}
  \caption{The ratio between the analyzing power $<A_z^{th}>$  
    calculated with grids with smaller numbers of grid points and the analyzing 
    power $<A_z^{th}>^{big}$ obtained with the final, fine grid. The ratio of the two 
    values is shown as a point for each of the 36 bins of the independent variable $\Delta\phi$.}
  \label{fig:test-grid}
\end{figure}

The required mesh size has been 
determined such that the shape and magnitude of $<O^{th}>$ is sufficiently
insensitive to a change of the number of grid points.
In order to test how the interpolated values depend 
on the mesh size of the grid, we performed calculations
 using grids with different 
numbers of points. The average 
of Eq.~(\ref{eq:method-6}) is always taken over the same 
list of about $3*10^6$ phase space 
coordinates, culled from an actual experiment~\cite{IUCFexp}. In 
Fig.~\ref{fig:test-grid} we compare the analyzing power $<A_z^{th}>$  
calculated with grids of fewer points (coarser mesh size)
 to the corresponding 
$<A_z^{th}>^{big}$ obtained with the fine grid described above. 
The ratio of the two 
values is shown as a point for each of the 36 bins of the independent 
variable $\Delta\phi$. 
A vertical spread of points is thus 
a measure for the interpolation uncertainty.

\begin{figure}
  \centering
  \includegraphics[width=10cm,bb=150 30 670 550,clip=true]{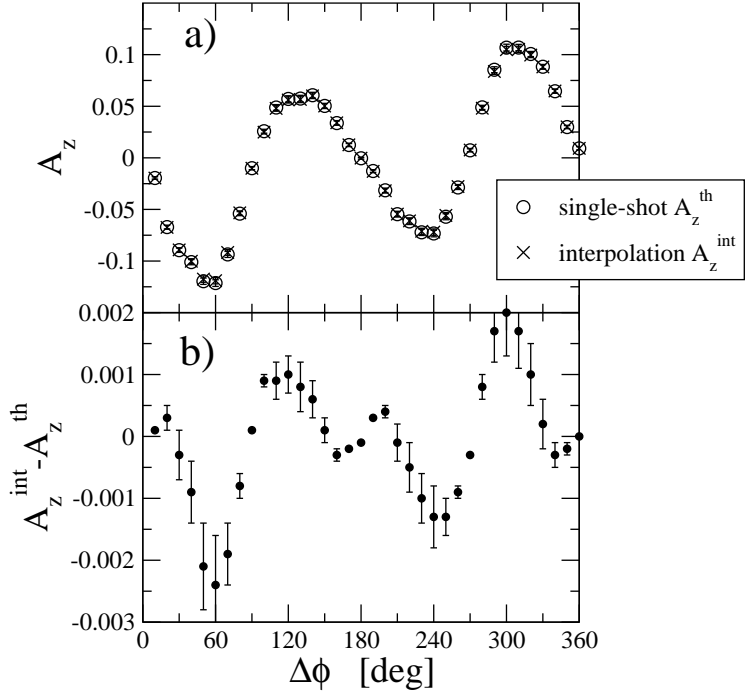}
  \caption{a) comparison of the single-shot calculation of 
    $A^{th}_z$ (open circles) with the grid interpolation 
    method $A^{int}_z$ (crosses); b) the difference between the two calculations}
  \label{fig:pia-joanna}
\end{figure}

To compare the (time-consuming) single-shot mode with the (fast) 
grid interpolation 
method, we have calculated the analyzing power $<A_z^{th}>$ as a function 
of $\Delta \phi$ using both methods. The values of $A^{th}_z$ for 
all events within 
a given $\Delta \phi$ bin are averaged according to Eq.~(\ref{eq:method-6}) 
to give 
$<A_z>$ for that bin.  
The error bars are calculated 
according to Eq.~(\ref{eq:method-7}) and turn out to be smaller than 0.003.
The results from the single-shot mode (circles) and the 
grid interpolation method (crosses) are shown in Fig.~\ref{fig:pia-joanna}a.
The difference between the two results is shown 
in Fig.~\ref{fig:pia-joanna}b. Again, this difference is a measure  of the 
interpolation uncertainty, or, in other words, of the 'curvature' of the 
interpolated function (hence the systematic dependence on $\Delta\phi$).   

\section{EXAMPLES}
\label{sec:examples}

Recently, an experimental study of the $dp$ break-up reaction with polarized 
270 MeV deuterons on a polarized proton target has been completed using the 
``Cooler'' storage ring at 
the Indiana University Cyclotron Facility~\cite{IUCFexp}. 
In this section, we illustrate the use of our new method 
to obtain theoretical estimates of polarization observables 
for this experimental setup. 

We calculated the differential cross section and  several polarization 
observables $O^{th}(x)$ 
using the CD Bonn NN potential with and without the TM' 3NF 
for  each  grid point in the available phase space.
After that, using the multidimensional interpolation, cf. the appendix, 
we obtained theoretical predictions for each experimental phase space point 
(about $3*10^6$ experimental points). 
Using the averaging method of Eq.~(\ref{eq:method-6}), the boundaries 
of the experimental acceptance for the energies and azimuths of 
the two outgoing protons, which  are approximately 
$E_1^{lab}, E_2^{lab} > 50$ MeV, 
$10^\circ \le \theta_{1}^{lab}, \theta_{2}^{lab} \le 45^\circ $, 
respectively, were automatically taken into account. The errors from the 
sampling statistics are obtained from Eq.~(\ref{eq:method-7}).

\begin{figure}
  \centering
  \includegraphics[width=10cm,bb=0 30 720 530,clip=true]{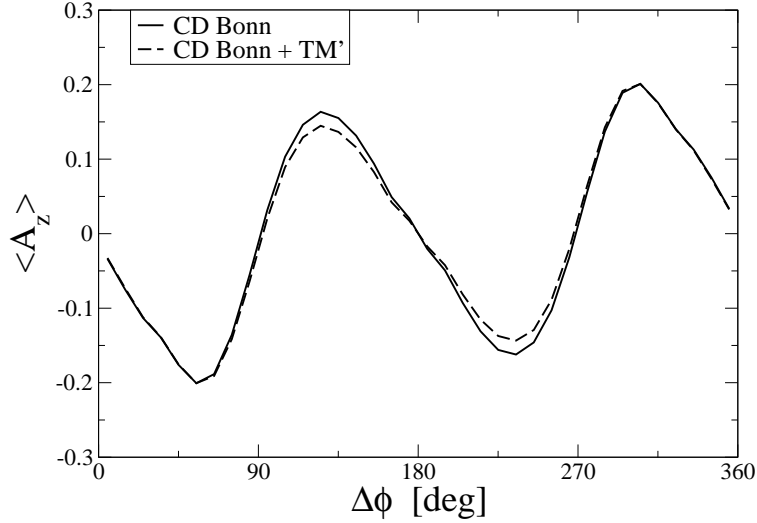}
\caption{Average value of vector analyzing power as a function of 
$\Delta \phi$. The solid and dashed lines are the CD Bonn and CD Bonn + 
TM' 3NF predictions, respectively.}
\label{fig:az-phinn}
\end{figure}

\begin{figure}
  \centering
  \includegraphics[width=10cm,bb=0 30 720 530,clip=true]{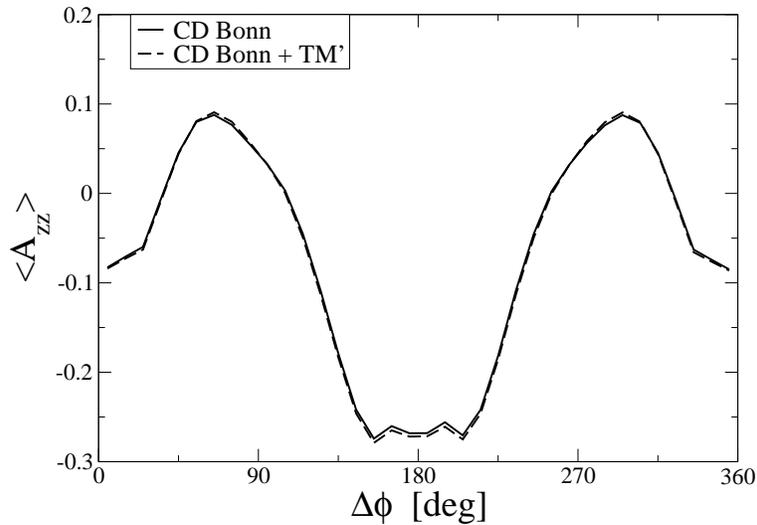}
  \caption{Average values of tensor analyzing power as 
    a function of $\Delta \phi $. The solid and dashed lines 
    are the CD Bonn and CD Bonn + 
    TM' 3NF predictions, respectively.}
\label{fig:tensor-phinn}
\end{figure}

\begin{figure}
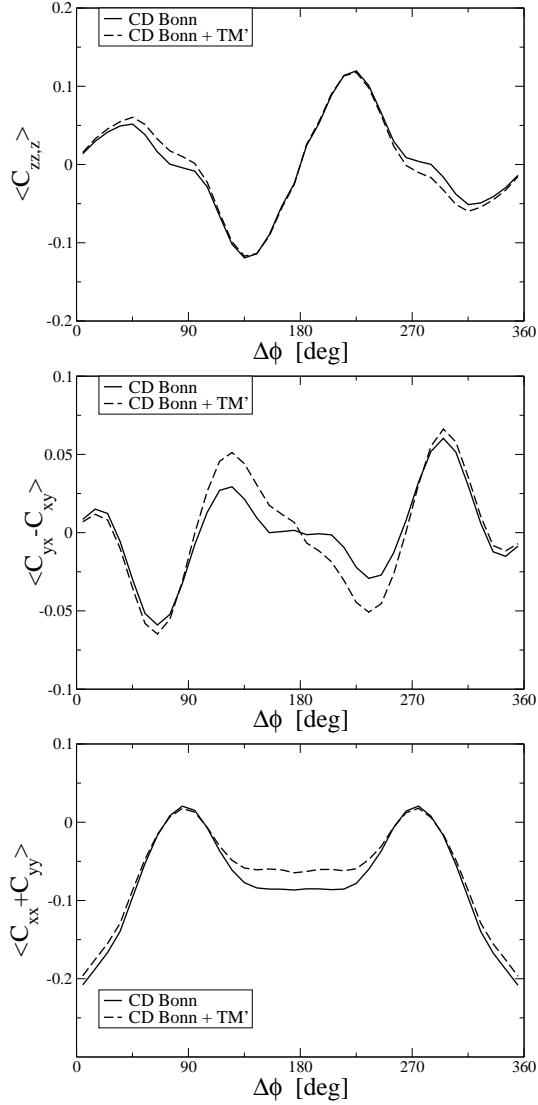

  \centering
  \includegraphics[width=7cm,bb=0 30 720 530,clip=true]{fig6.1.eps}\\
  \includegraphics[width=7cm,bb=0 30 720 530,clip=true]{fig6.2.eps}\\
  \includegraphics[width=7cm,bb=0 30 720 530,clip=true]{fig6.3.eps}
  \caption{Average values of spin correlation coefficients 
    as a function of $\Delta \phi $.  
    The solid and dashed lines are the CD Bonn and CD Bonn + 
    TM' 3NF predictions, respectively.}
  \label{fig:corefi-phinn}
\end{figure}

The results  for the vector analyzing power $A_z$, the tensor analyzing power 
 $A_{zz}$ and the correlation coefficients $C_{zz,z}$, $C_{yx}-C_{xy}$
and $C_{xx}+C_{yy}$  are shown in  
Figs~(\ref{fig:az-phinn}) -~(\ref{fig:corefi-phinn}) 
as a function of $\Delta \phi$. In these figures the solid line
shows the Faddeev calculations based on the CD Bonn NN potential, while the 
dashed line is obtained by including  the TM' 3NF.   
The errors  (less than 0.003) are  too small to be visible in figures.

The difference between the solid and the dashed lines
illustrates the effect of the TM' 3NF.
As one can see, the 3NF effects are rather small in the 
present case where the observables are integrated over a large portion of the 
phase space. However, 
in the case of some  correlation coefficients  
(Fig.~\ref{fig:corefi-phinn}) the 
magnitude of 3NF effects is sufficiently large 
to be distinguishable by a future 
experiment. 


It is conceivable that the smallness of the 3NF effects in some of the cases 
that are shown 
here is due to the fact the observable is averaged over a large portion of 
the phase space.
In order to identify regions in phase space with strong 3NF effects, we have 
searched 
for the largest differences between observables predicted with 
 the CD Bonn interaction either by itself, or with the TM' 3NF included. 
The influence of the 3NF can be quantified by a difference 
$\Delta O^{th}$ at a given phase space point when the  3NF is switched on
\[
\Delta O^{th} =|O^{th}(2N)-O^{th}(3N)|
\]  
Here $O^{th}(2N)$, $O^{th}(3N)$ are predictions for the particular 
observable resulting when using the CD Bonn potential alone and 
including in addition the  TM' 3NF, respectively. 

First, for each observable $O^{th}$ we searched over the entire grid 
of the phase space points for the maximum value 
of the difference $\Delta O^{th} $. 
The results are presented in Table~I. 
Second, since we are looking  for the effects that should 
be experimentally accessible, we searched for the 
maximum of $\Delta O^{th} $  only through the phase 
space points for which the differential cross section 
$\frac{d\sigma }{d\hat p d\hat q dq} = 
\sigma (2N) > 0.001 \frac{fm^2}{MeV sr^2}$. 
The resulting 
$max [\Delta O^{th}]$ are shown in third column in Table~I.
In addition, in the brackets, labeled (*) and (**) of Table~I, we 
 list the number of grid points for which 
$0.05 \le \Delta  O^{th} \le  max [\Delta  O^{th}]$ and 
$0.05 \le \Delta  O^{th} \le  max [\Delta  O^{th}]$ 
with  $\sigma (2N) > 0.001\frac{fm^2}{MeV sr^2}$, respectively.
\begin{table}[hbpt]
\label{table}
\caption{Maximum value of difference 
$\Delta O^{th} = |O^{th}(2N)-O^{th}(3N)|$  for each observable $O^{th}$.
The second column, labeled $max [\Delta O^{th}]$, 
gives results without any restrictions. 
The third one, labeled $max [\Delta O^{th}] + \sigma$, 
results after applying cross section limitation 
( $\sigma (2N) > 0.001 \frac{fm^2}{MeV sr^2}$).
In  the brackets, in the column labeled (*), 
there is a number of grid points for which 
$0.05 \le \Delta O^{th} \le  max [\Delta  O^{th}] $. 
In the column labeled (**) there are  number of grid points
   for which $0.05 \le \Delta O^{th} \le  max [\Delta  O^{th}] $ 
and $\sigma (2N) > 0.001 \frac{fm^2}{MeV sr^2}$.}
  \begin{center}
    \begin{tabular}{|c|lr|lr|}
      \hline
    $O^{th}$ & $max [\Delta O^{th}]$ &(*)& $max [\Delta O^{th}] + \sigma$ &(**) \\
             
      \hline
      \hline
      $A_z$ &0.05&(10)&&(0)\\ 
      \hline
      $ A_{zz}$       &0.12 &(62 150)&0.12&(11 409)\\
      \hline
      $C_{xx}+C_{yy}$    &0.18 & (70 110)&0.11&(1 458)\\
      $C_{yx}-C_{xy}$    &0.08 & ( 4 350)&    &(0)\\
      $C_{zz,z}$         &0.10 & ( 2 376)&    &(0)\\  
      \hline
    \end{tabular}
  \end{center}
\end{table}
As we can see in Table~I the tensor analyzing power $ A_{zz}$ or
the correlation coefficient $C_{xx}+C_{yy}$ 
are especially interesting, showing large 3NF effects, on the other hand 
the vector analyzing power $A_z$ 
shows the least  evidence of the action of the TM' 3NF. 

\begin{figure}[htbp]
  \centering
  \includegraphics[width=10cm,bb=160 290 570 550,clip=true]{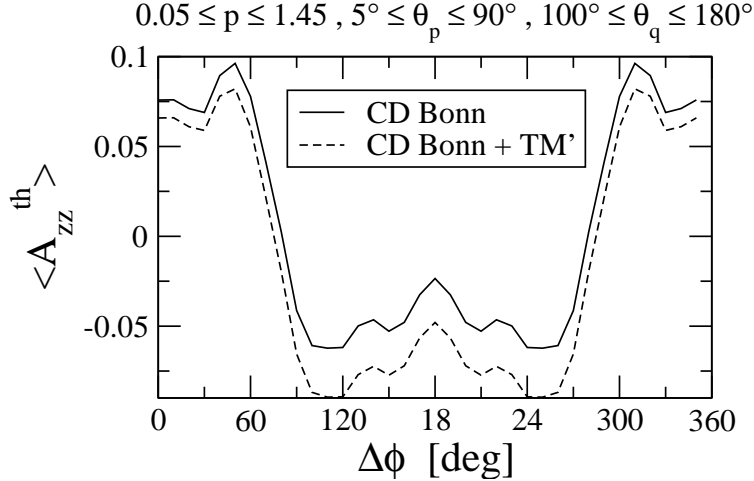}
  \caption{The average value of tensor  analyzing power $A_{zz}$ 
    as a function of $\Delta \phi $, in a region 
    of phase space that is sensitive to the inclusion of a three-nucleon force 
    ($0.05 \le p \le 1.45, 5^\circ \le \theta_p \le  90^\circ, 
    100^\circ \le \theta_q \le 180^\circ $). 
    The solid and dashed lines are the CD Bonn and CD Bonn + 
    TM' 3NF predictions, respectively.}
  \label{fig:eg-azz}
\end{figure}

The  next step in our study was to find the regions of the phase space with 
the largest value of $\Delta O^{th}$. 
Let us investigate here, only for example, the dependence of 
$A_{zz}(\Delta \phi)$ on the additional variables 
$p$, $\theta_p$ and $\theta_q$. 
For this purpose, the four-dimensional phase space distributions of 
$<A_{zz}^{th}>$  and  $\Delta A_{zz}^{th}$
are projected onto three planes: $\Delta \phi - p$, $\Delta \phi - \theta_p$ 
and  $\Delta \phi - \theta_q$. In the case of 
$\Delta A_{zz}^{th}$ for each projection 
point the maximum value over the range of the remaining variables is 
determined,  e.g. the $\Delta \phi - p$ projection shows 
$max [ \Delta A_{zz}^{th} 
(p,\theta_p,\theta_q,\Delta \phi)]_{ \Delta \phi, p}$. 
\begin{figure}
  \centering
  \includegraphics[width=13cm,bb=0 0 570 570,clip=true]{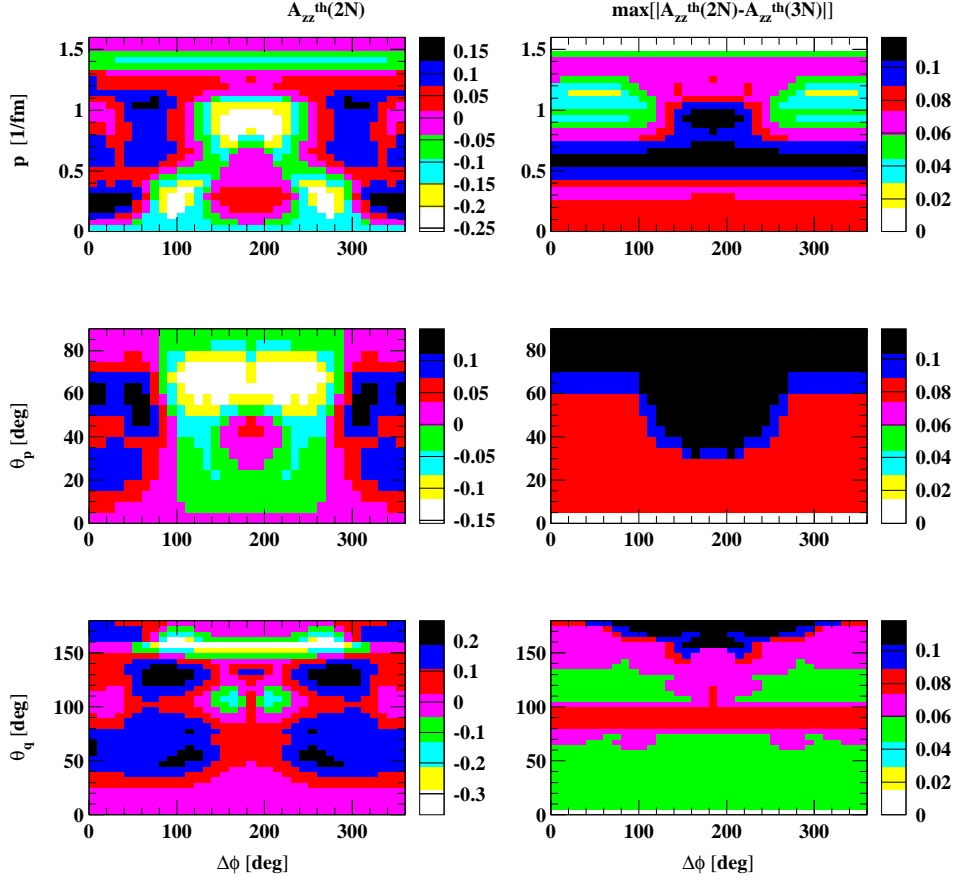}
  \caption{Two-dimensional projections of theoretical prediction 
    for tensor analyzing power $A_{zz}^{th}$. Columns show the average value 
    of $<A_{zz}^{th}(2N)>$
    and  the maximum of difference $|A_{zz}^{th}(2N)-A_{zz}^{th}(3N)|$.
    The three rows show projections onto the 
    $\Delta \phi - p$, $\Delta \phi - \theta_p$ 
    and  $\Delta \phi - \theta_q$ planes, respectively.}
  \label{fig:map-azz}
\end{figure}
Using graphs from the second column we can see which region of phase space 
exhibits a large 3NF effect. The figure shows, for example, that 
an interesting region would be $100^\circ \le \theta_q \le 180^\circ$
with the full range of $p$ and $\theta_p$ included. Fig.~\ref{fig:eg-azz} 
shows the average value of $<A_{zz}^{th}>$ 
after integration over all values of 
$p$, $\theta_p$ and integration over part 
of $100^\circ \le \theta_q \le 180^\circ$. For this example the 3NF effects 
are quite large.

Studies such as this, together with earlier work published in ~\cite{kuros03},
  may be used to identify regions of 
enhanced sensitivity the the 3NF in order to plan future experiments.

\section{ SUMMARY}
\label{sec:summary}
We have presented a simple, straight-forward method to integrate over regions
of phase space the prediction from a theoretical model. 
The method reflects the 
true experimental acceptance, as well as variations of the detector efficiency.
It is applicable to kinematically complete experiments with three 
bodies in the final state, and eliminates the need for 
a Monte-Carlo simulation 
of the detector setup.

We have applied this method to several observables of $dp$ 
break-up at 270 MeV, and studied
  the effect of including a three-nucleon force 
in the model.

\acknowledgments

The work has been supported by the U.S. National Science Foundation 
under Grant NO. PHY-0100348, the Swedish Research Council Dnr 629-2001-3868, 
DOE grants DE-FC02-01ER41187, DE-FG03-00ER41132  
and by the Polish Committee for Scientific Research under Grant No.
2P03B00825.
The authors would like thank to Professor W. Gloeckle for fruitful 
discussions and remarks.
One of us (R.S.) would like to  thank
the Foundation for Polish Science for a financial support.
The Faddeev calculations have been performed on the SV1 and
the CRAY T3E of the NIC in Juelich, Germany.

\appendix
\section{Multidimensional linear interpolation}
The sampling method described in section~\ref{sec:method}, requires that the
theoretical value of a given observable is calculated for each of a large
number of events, collected in an experiment. In order to achieve the 
necessary processing speed, precalculated theoretical values are stored in a 
four-dimensional matrix (the 'grid') as explained in 
Sec.~\ref{sec:Computational details}. Multidimensional
 linear interpolation is then 
used to retrieve the corresponding value at any phase space point.

We denote the set of four continuous independent variables as
$x = \{ p, \theta_p, \theta_q, \Delta\phi \} \equiv \{ x_i\}$; $(i=1..4)$. 
The corresponding {\it discrete} independent variables 
that constitute the grid 
are $x_{i,k_i}$, $k_i=1..n$, where $n$ is the size of the grid in the $i$th 
dimension. It is understood that $x_{i,k_i} \leq x_i \leq x_{i,k_i+1}$
for any value $x_i$ of the set $x$ at which a functional 
value is sought, i.e. the elements of the discrete grid 
$x_{i,k_i}$-variables are 
strictly increasing and the $x_i$ coordinate at which interpolation is 
performed lies always inside an interval of grid points.
Applying linear interpolation~\cite{bib:LinInt} in four dimensions, means that 
the grid cell that contains the argument $x$ of interest has $2^4=16$ corners.
We are using the FORTRAN routine FINT~\cite{bib:FINT} from the CERN program
library. The interpolation is carried 
out as follows.

The interpolation weights, $w_i$, are defined as 
\begin{equation}
w_i=\frac{x_i - x_{i,k_i}}{x_{i,k_i+1} - x_{i,k_i}}
\end{equation}
For brevity we also define $u_i = 1 - w_i$. The functional values at the corners
of the grid cell that contains $x$ are denoted by $y_m$; $m=1..16$, and defined 
as follows, 
\begin{equation}
\begin{array}{l}
y_1 \equiv O(x_{1,k_i},x_{2,k_i},x_{3,k_i},x_{4,k_i})\\
y_2 \equiv O(x_{1,k_i+1},x_{2,k_i},x_{3,k_i},x_{4,k_i})\\
y_3 \equiv O(x_{1,k_i+1},x_{2,k_i+1},x_{3,k_i},x_{4,k_i})\\
y_4 \equiv O(x_{1,k_i},x_{2,k_i+1},x_{3,k_i},x_{4,k_i})\\
y_5 \equiv O(x_{1,k_i},x_{2,k_i},x_{3,k_i+1},x_{4,k_i})\\
y_6 \equiv O(x_{1,k_i+1},x_{2,k_i},x_{3,k_i+1},x_{4,k_i})\\
y_7 \equiv O(x_{1,k_i+1},x_{2,k_i+1},x_{3,k_i+1},x_{4,k_i})\\
y_8 \equiv O(x_{1,k_i},x_{2,k_i+1},x_{3,k_i+1},x_{4,k_i})\\
......\\
y_{16} \equiv O(x_{1,k_i},x_{2,k_i+1},x_{3,k_i+1},x_{4,k_i+1})\\
\end{array}
\end{equation}
The observable at any point $x=(x_1,x_2,x_3,x_4,x_5)$ is then given by
\begin{equation}
\begin{array}{l}
O(x_1,x_2,x_3,x_4) = \\
u_1 u_2 u_3 u_4 \cdot y_1 +    w_1 u_2 u_3 u_4 \cdot y_2 +
w_1 w_2 u_3 u_4 \cdot y_3 +    u_1 w_2 u_3 u_4 \cdot y_4 + \\
u_1 u_2 w_3 u_4 \cdot y_5 +    w_1 u_2 w_3 u_4 \cdot y_6 + 
w_1 w_2 w_3 u_4 \cdot y_7 +    u_1 w_2 w_3 u_4 \cdot y_8 + \\
u_1 u_2 u_3 w_4 \cdot y_9 +    w_1 u_2 u_3 w_4 \cdot y_{10} +  
w_1 w_2 u_3 w_4 \cdot y_{11} + u_1 w_2 u_3 w_4 \cdot y_{12} + \\
u_1 u_2 w_3 w_4 \cdot y_{13} + w_1 u_2 w_3 w_4 \cdot y_{14} +  
w_1 w_2 w_3 w_4 \cdot y_{15}+ u_1 w_2 w_3 w_4 \cdot y_{16}
\end{array}
\end{equation}

\begin{figure}
[htbp]
  \centering
  \includegraphics[width=10cm,bb=0 45 715 530,clip=true]{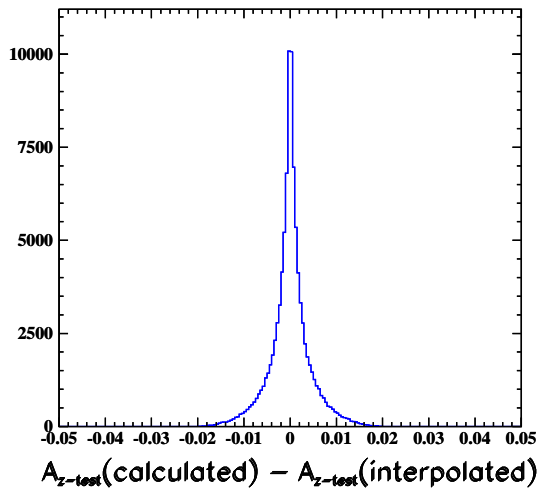}
\caption{The difference between the interpolated and the exact test observable.} 
\label{fig:DAZFAKE}
\end{figure} 

In order to investigate the accuracy of the interpolation routine, we
construct a 'test' grid that is analogous to the one described in 
Sec.~\ref{sec:Computational details}, but with function values that are 
given by the analytic expression
\begin{equation}
\begin{array}{l}
O(x) = A_{z-test}(p,\theta_p,\theta_q,\phi_p)= \\
4{p}/{p_{max}}(1-{p}/{p_{max}}) \cdot 
(\sin(\theta_q) \sin(2\theta_p) \sin(\phi_p) +
\sin2(\theta_q) \sin2(\theta_p) \sin(2\phi_p))
\end{array}
\end{equation}
where  $p_{max}$ is the largest possible relative proton momentum. 
This function has been chosen to represent a possible, 
typical dependence of $A_z$ on 
the final-state kinematics.
 
For a large number of arguments $x$, corresponding to random, 
phase-space-distributed break-up events, the exact value of the function is 
compared to the value returned by the interpolation routine, acting on the grid
values. The distribution of differences is shown in Fig.~\ref{fig:DAZFAKE}. 
The 
spread of values around zero arises because the interpolation assumes that the 
function is linear within a grid cell. This effect causes a systematic shift 
between the interpolated and the true values. For this reason, the FWHM of 
0.003 observed in Fig.~\ref{fig:DAZFAKE} should be compared to the value of 
the function. When averaging over regions of phase space this effect 
washes out 
further, becoming clearly insignificant  indicating  that the mesh size 
of the grid
used here is sufficiently small.

\end{document}